\documentclass[twocolumn]{aastex7}

\usepackage{color} 

\definecolor{mygreen}{RGB}{0, 185, 118}

\hypersetup{linkcolor=blue,citecolor=blue,urlcolor=blue}

\begin{document}

%
%


\title{Characterizing the ‘Martin Luther King’ Event: \\ I.  Sun-to-Earth Evolution of the January 19, 2026 Geomagnetic Storm }

\author[orcid=0000-0001-5400-2800]{Abril Sahade} 
\affiliation{Heliophysics Science Division, NASA Goddard Space Flight Center, Greenbelt, MD 20770, USA}
\email[show]{abril.sahade@nasa.gov}

\author[orcid=0009-0009-9799-979X]{Cecilia Mac Cormack}
\affiliation{Heliophysics Science Division, NASA Goddard Space Flight Center, Greenbelt, MD 20770, USA}
\affiliation{The Catholic University of America, Washington, DC 20064, USA}
\email{cecilia.maccormack@nasa.gov}  

\author[orcid=0000-0002-6273-4320]{Andreas J. Weiss} 
\affiliation{Heliophysics Science Division, NASA Goddard Space Flight Center, Greenbelt, MD 20770, USA}
\affiliation{Goddard Planetary Heliophysics Institute, University of Maryland, Baltimore County, Baltimore, MD 21250, USA}
\email{andreas.j.weiss@nasa.gov}

\author[orcid=0000-0002-3746-9246]{Antonio Niemela} 
\affiliation{Heliophysics Science Division, NASA Goddard Space Flight Center, Greenbelt, MD 20770, USA}
\affiliation{Goddard Planetary Heliophysics Institute, University of Maryland, Baltimore County, Baltimore, MD 21250, USA}
\email{antonio.e.niemelaceleda@nasa.gov}

\author[orcid=0000-0003-1402-0398]{Diego G. Lloveras} 
\affiliation{Heliophysics Science Division, NASA Goddard Space Flight Center, Greenbelt, MD 20770, USA}
\affiliation{George Mason University, Fairfax, VA, 22030, USA}
\email{diego.lloveras@nasa.gov}

\author[orcid=0000-0001-6590-3479]{Erika Palmerio} 
\affiliation{Predictive Science Inc., San Diego, CA 92121, USA}
\email{epalmerio@predsci.com}

\author[orcid=0000-0003-0565-4890]{Teresa Nieves-Chinchilla}
\affiliation{Heliophysics Science Division, NASA Goddard Space Flight Center, Greenbelt, MD 20770, USA}
\email{teresa.nieves@nasa.gov}

\begin{abstract}
We present a comprehensive Sun-to-Earth analysis of the extreme space weather event of 18–-19 January 2026, hereafter the Martin Luther King (MLK) storm, which produced a severe (G4) geomagnetic storm and one of the most intense solar radiation storms of Solar Cycle 25. The geomagnetic disturbance was driven by a fast Earth-directed coronal mass ejection (CME) associated with an X1.9-class flare from NOAA Active Region 14341. We combine multi-viewpoint remote-sensing observations, in situ measurements from Solar Orbiter and Wind, and physics-based magnetohydrodynamic simulations using the CORHEL-CME framework to reconstruct the eruption from the low corona to 1 au and identify the mechanisms governing its geoeffectiveness.
Observations and simulations indicate that the eruption originated from a right-handed magnetic flux rope embedded in a complex magnetic environment dominated by a neighboring positive-polarity coronal hole. The simulations reproduce the observed CME morphology, asymmetric expansion, magnetic fields measured at both spacecraft, and Dst index evolution. The main discrepancy is attributed to filament material not included in the model. Favorable magnetic polarity alignment between the CME and coronal-hole field minimized magnetic erosion through interchange reconnection, enabling rapid escape and preservation of the strong southward magnetic field that drove the geomagnetic storm. The prolonged recovery phase resulted from interaction between the trailing CME and a following high-speed stream from the same coronal hole. These results demonstrate that storm intensity was primarily established by the coronal magnetic configuration, whereas storm duration was controlled by the ambient solar wind.
\end{abstract}


\section{Introduction}

Extreme solar events represent some of the most energetic phenomena in our solar system, involving the rapid release of large amounts of stored magnetic energy into the heliosphere. This energy may be emitted as electromagnetic radiation in the form of powerful X-class solar flares \citep[][]{fletcher_2011}, transferred as kinetic and magnetic energy through coronal mass ejections \citep[CMEs;][]{chen_2011,webb_2012}, and used to accelerate particles to high energies, both in flare reconnection sites and at CME-driven shocks. Those accelerated particles that escape into interplanetary space and are detected in situ are known as solar energetic particles \citep[SEPs;][]{reames_1999,reames_2013}.  
These processes are often coupled, producing far more severe impacts on the near-Earth space environment than would be expected from any single mechanism acting alone \citep{gopalswamy_2005}. 

CMEs represent the most significant drivers of geomagnetic activity, as their large-scale magnetic structures interact directly with Earth's magnetosphere during transit through the heliosphere. The intensity and duration of a geomagnetic storm strongly depend on the magnetic configuration of the solar wind structure interacting with the magnetosphere. In fact, two of the key factors controlling geomagnetic storm intensity are the magnitude and orientation of the IMF north--south ($B_z$) component \citep{1999Gonzalez,2006Schwenn}. When $B_z$ remains negative for an extended period, magnetic reconnection at the dayside magnetopause is enhanced, allowing efficient energy transfer from the solar wind into Earth's magnetosphere and leading to significant geomagnetic activity \citep{kamide_1998,1999Daglis,2006Tsurutani}. The combination of high SEP fluences and intense geomagnetic activity poses significant hazards to satellites, power grids, aviation, and human spaceflight operations \citep{eastwood_2017,2021Singh}, motivating continued advances in both the scientific understanding and operational forecasting of such events \citep{2015Schrijver,2023Temmer}.

On January 19, 2026, Earth was impacted by a major space weather event associated with a fast, Earth-directed CME and its driven shock. The eruption, which occurred on January 18, 2026, was accompanied by a long-duration X1.9-class solar flare and produced a rapid enhancement of high-energy particles, resulting in a severe (S4) solar radiation storm as determined by the proton flux measured near Earth. The CME arrived at Earth at approximately 19:00 UT on January 19, driving a geomagnetic storm that reached G4 (severe) levels on the NOAA Space Weather Scales\footnote{See \url{https://www.spaceweather.gov/noaa-scales-explanation} for a detailed explanation of the S and G storm scales.}. This event, hereafter referred to as the ‘MLK\footnote{Named in recognition of its occurrence on the Martin Luther King Jr.\ Day, a U.S.\ federal holiday.} storm,’ represents a benchmark case study for multi-disciplinary space weather research. The MLK storm combined both extreme particle acceleration and intense geomagnetic forcing, creating a rare observational opportunity to study the complex coupling between coronal dynamics, heliospheric propagation, and near-Earth space environment response. Such complex events, in which SEP enhancements and strong geomagnetic storms occur at the same time, are statistically infrequent but mostly important for both scientific characterization of extreme space weather and for the validation of operational forecasting models \citep{riley_2012,schrijver_2012}.

This paper presents a comprehensive analysis of the MLK storm to understand the chain of physical processes linking the solar eruption to the resulting G4 geomagnetic storm at Earth. We first establish the key physical properties that can be robustly inferred from remote-sensing and in situ observations, with particular emphasis on connecting coronal observations to the magnetic structure measured near 1~au. To bridge the observational gap between the solar corona and Earth, we employ physics-based simulations using the CORHEL-CME framework \citep{linker_2024}. The simulations serve two complementary purposes: (1) to reconstruct the three-dimensional evolution of the CME and its interaction with the ambient solar wind, thereby providing a continuous description of its propagation through the heliosphere, and (2) to test and refine hypotheses derived from the observations regarding shock formation, particle acceleration, and the magnetic field orientation responsible for the geomagnetic storm. This analysis is performed iteratively, with observational constraints informing the simulation setup and the simulation results subsequently guiding the interpretation of the observations. Such an integrated approach provides a more self-consistent characterization of the event than can be achieved using observations or modeling alone, yielding new insight into the physical processes governing extreme Sun-to-Earth space weather events \citep[e.g.,][]{maccormack_2025b,sahade_2026}.

While the present paper focuses on the Sun-to-Earth evolution of the CME and the origin of the associated geomagnetic storm, the solar energetic particle event responsible for the severe (S4) radiation storm will be examined in a companion paper.
The paper is organized as follows. Section~\ref{Obs} presents the observational data sets and provides an overview of the event. Section~\ref{mod} describes the CORHEL-CME simulation setup. Section~\ref{res} presents the observational and modeling results, and Section~\ref{conc} summarizes the main conclusions and discusses their implications for future space weather forecasting.

\section{Observations}\label{Obs}
\subsection{Source region and coronal evolution}\label{SR}

\begin{figure*}
    \centering
    \includegraphics[width=\linewidth]{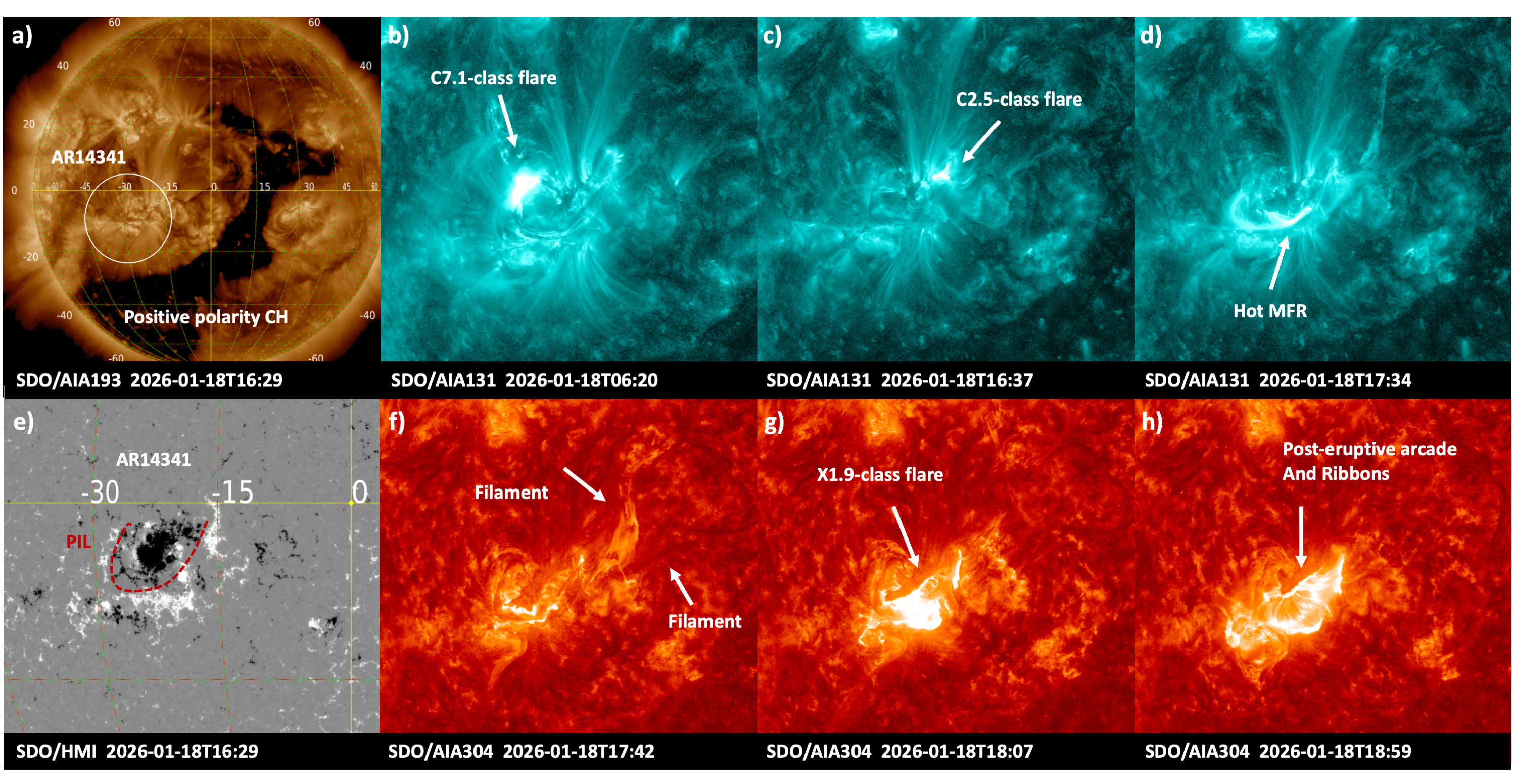}
    \caption{SDO/AIA and HMI observations of the Sun and AR14341. \textbf{(a)} An overview of the solar disk with AR14341 and the surrounding coronal hole observed with SDO/AIA 193$\,$\AA{} on January 18 at 16:29$\,$UT. \textbf{(b, c, d)} AR14341 with the C7.1-class and C2.5-class flare, and a hot magnetic structure, all observed with SDO/AIA 131$\,$\AA{} at 06:20$\,$UT, 16:37$\,$UT and 17:34$\,$UT, respectively. \textbf{(e)} AR14341 HMI line-of-sight magnetogram with a red dashed curve drawn over the U-shaped PIL. \textbf{(f)} Quiescent filaments near AR14341 that interact with the main eruptive structure. \textbf{(g, h)} A X1.9-class flare and L-shaped two-ribbon configuration and post-eruptive arcade observed with SDO/AIA 131$\,$\AA{} at 18:07$\,$UT and 18:59$\,$UT, respectively.}
    \label{fig:sr}
\end{figure*}

\begin{figure*}
    \centering
    \includegraphics[width=0.9\linewidth]{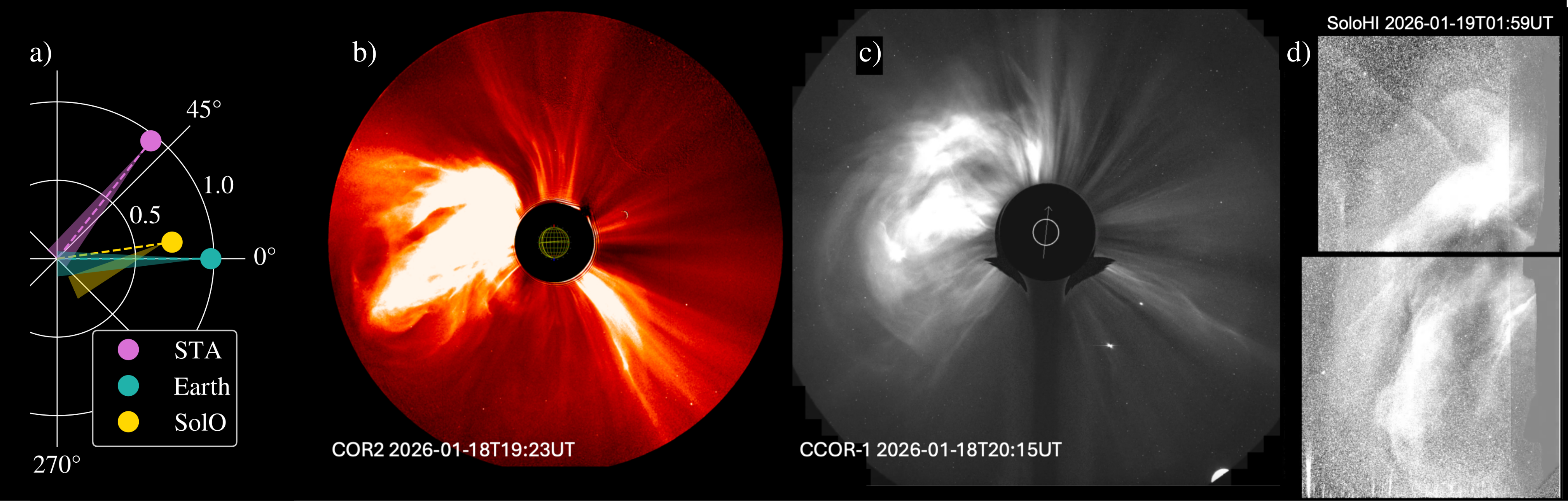}
    \caption{Remote white-light observations of the MLK event from different spacecraft and viewpoints. \textbf{(a)} Spacecraft locations and field of views. \textbf{(b)} STEREO-A COR2 observation for the MLK event at 19:23$\,$UT. \textbf{(c)} CCOR-1 observations of the event at 20:15$\,$UT. \textbf{(d)} SoloHI inner FOV observation on January 19 at 02:2301$\,$UT processed with base difference. An animated version of this figure shows STEREO-A COR2, followed by CCOR-1 observations, from January 18 17:00 UT to January 19 03:00 UT, and SoloHI observations from January 18 19:00 UT to January 20 00:00 UT.}
    \label{fig:cor}
\end{figure*}

The MLK storm was driven by a CME originating on January 18, 2026 from the $\beta\gamma$ active region (AR) 14341, located at heliographic coordinates S10E20. The active region and its evolution can be followed in multiple wavelengths from the Earth perspective. Figure~\ref{fig:sr}a illustrates, in particular, the location of the AR and the surrounding coronal hole as observed in the 193$\,$\AA{} wavelength of the Atmospheric Imaging Assembly \citep[AIA;][]{lemen2012atmospheric} on board the Solar Dynamics Observatory \citep[SDO;][]{pesnell_2012}. The region exhibited a compact, nearly circular magnetic configuration indicative of a highly concentrated magnetic flux distribution. It hosted a well-defined U-shaped filament channel tracing the main polarity inversion line (PIL), as shown in a line-of-sight magnetogram from the Helioseismic and Magnetic Imager \citep[HMI;][]{scherrer_2012} instrument, onboard SDO, in Figure~\ref{fig:sr}e. 

EUV and photospheric magnetic field observations together allow us to infer the handedness of the erupting structure \citep[e.g.,][]{2017Palmerio}. The AIA observations of the activated hot magnetic structure show its positive and negative footpoints anchored near Carrington coordinates (50$^{\circ}$, $-8^{\circ}$) and (40$^{\circ}$, $-10^{\circ}$), respectively. The comparison of these locations with the orientation of the PIL and the polarity of the surrounding field indicates a sinistral configuration \citep[see, e.g., Figure$\,$2 of][]{2020Zhou}. Since sinistral filament channels are associated with right-handed flux ropes, we conclude that the main eruptive structure was a right-handed (positive helicity) flux rope.

At 06:20$\,$UT, the eastern segment of the filament channel produced a C7.1-class flare associated with a partially confined and relatively minor eruption (Figure~\ref{fig:sr}b). Although no CME was launched at this stage, this event likely played a critical role in the reconfiguration of magnetic free energy in the corona. Previous studies \citep[e.g.,][]{nindos_2020,kliem_2021} have shown that confined eruptions can facilitate the formation, or energization, of a temporarily stable magnetic flux rope (MFR). In this context, the early C7.1-class flare may have increased the axial flux and twist of an existing MFR, pushing the system closer to the eruptive threshold while remaining temporarily in equilibrium. The present case appears consistent with this mechanism and further illustrates the potential importance of precursor confined eruptions in assessing CME initiation and eruption likelihood \citep[as also discussed in][]{2020Patsourakos}.

At 16:37$\,$UT, a C2.5-class flare occurred near the western portion of the same filament channel (Figure~\ref{fig:sr}c). This event appears to have destabilized the pre-existing hot MFR, triggering its full eruption. The erupting MFR rapidly accelerated after $\sim$ 17:30$\,$UT, producing a fast CME. Magnetic reconnection beneath the erupting structure intensified dramatically, leading to the onset of a major X1.9-class flare at 18:07$\,$UT. Flare ribbons developed to the west (18:20 UT) and one hour later to the east, producing an L-shaped two-ribbon configuration and post-eruptive arcade, as shown in Figure~\ref{fig:sr}g and h. This sequence suggests a two-stage evolution in which early confined activity primed the system, while subsequent reconnection triggered the catastrophic loss of equilibrium \citep[see also][]{2020James}.

During its eruption, the main MFR interacted with nearby quiescent filaments (Figure~\ref{fig:sr}f). These structures were destabilized and erupted simultaneously with the expansion of the CME volume. Their eruption likely contributed additional mass to the ejecta, producing an overdense component that is later identified in both coronagraph imagery and in situ solar wind measurements, as discussed below. Such mass loading can significantly affect CME propagation and magnetic cloud structure at 1$\,$au. In addition, the eruption encountered the open magnetic field associated with a western coronal hole of positive polarity, i.e.\ with the same sign as the western footpoint of the MFR. 
Owing to the matching magnetic polarities, the interaction between the expanding CME and the open coronal-hole field likely influenced the lateral expansion and deflection of the CME while reducing magnetic erosion through interchange reconnection.

The MLK CME was observed in white-light imaging from multiple vantage points by coronagraphs and heliospheric imagers located at different positions in the inner heliosphere. Figure~\ref{fig:cor} shows the spacecraft configuration (a) and white-light CME observations (b,c,d). An animated version of the evolving CME is available in the online version. The event was monitored near the Sun--Earth line by the Compact Coronagraph 1 \citep[CCOR-1;][]{thernisien_2025} coronagraph onboard the Geostationary Operational Environmental Satellite (GOES-19). CCOR-1 provides white-light coronagraph observations from the Sun--Earth line with a field of view extending from approximately 3.7 to 24~R$_{\odot}$. Additional observations of the low and middle corona were obtained by the Sun--Earth Connection Coronal and Heliospheric Investigation \citep[SECCHI;][]{howard2008sun} instrument suite onboard the Solar TErrestrial RElations Observatory \citep[STEREO;][]{kaiser2008stereo}, including the COR1 and COR2 coronagraphs. COR1 provides observations of the low corona from 1.4 to 4~R$_{\odot}$, while COR2 images the corona out to approximately 15~R$_{\odot}$. Finally, the large-scale heliospheric propagation of the eruption was tracked by the Solar Orbiter Heliospheric Imager \citep[SoloHI;][]{howard_2020} onboard the Solar Orbiter \citep[SolO;][]{muller_2020} mission. SoloHI observes a field of view of 40$^{\circ}$ of elongation starting at 5$^{\circ}$ off the eastern limb of the Sun. STEREO and SolO were located at $51.4^{\circ}$ and $8.1^{\circ}$ from the Sun--Earth line, respectively, as shown in Figure~\ref{fig:cor}a. This multi-viewpoint configuration allowed the CME to be tracked continuously from the low corona into the inner heliosphere.

From the STEREO-A perspective, the CME is first detected in observations from COR1 at approximately 17:46$\,$UT (not shown in figures). These observations capture the initial expansion of the eruptive structure in the low corona. At this stage, the CME exhibits a clear asymmetric expansion that becomes progressively more pronounced in subsequent COR2 observations (Figure \ref{fig:cor}b and animation from seconds 1 to 3). The asymmetry is characterized primarily by a restricted expansion along the lower portion of the CME front. This region appears to be bounded by a nearby western coronal hole, which likely acts as a magnetic barrier that limits the lateral expansion of the erupting structure and produces a flattened morphology in the southern flank of the CME. 

As the CME evolves in the COR2 field of view, its structure becomes more complex and two distinct components can be identified, as shown in Figure \ref{fig:cor}b. The primary body of the CME propagates slightly south of the solar equator, forming the dominant white-light structure of the eruption. A second brighter feature is visible in the northern portion of the ejecta. This northern density enhancement is most likely associated with filament material that erupted alongside the main MFR. The spatial separation between the main CME body and this overdense feature may reflect differences in the dynamics of the erupting filament system.

Figure~\ref{fig:cor}c shows the CME observed by CCOR-1, for which the event entered the field of view at approximately 18:30$\,$UT (see animation from seconds 3 to 7). In these observations, the leading edge of the disturbance is characterized by a rapidly expanding halo-like front consistent with the presence of a CME-driven shock. However, the bulk of the ejecta remains concentrated toward the eastern side of the field of view. Despite the different viewing geometry, the influence of the coronal hole on the lower portion of the CME remains evident, producing a flattened morphology along the southern flank. The filament-associated material is also visible in these observations, appearing once more in the northern region of the ejecta, although it is located more centrally with respect to the main CME structure. 

Finally, the CME enters the field of view of SoloHI at approximately 21:13$\,$UT (Figure \ref{fig:cor}d). The eastern portion of the CME front can be seen propagating through the first two tiles of the SoloHI field of view. The detection at these elongations indicates that the CME has transitioned from the coronagraph domain into the heliosphere while maintaining a coherent large-scale structure.

A clear shock signature associated with the CME can be identified in observations from all instruments considered here (see animated version of Figure~\ref{fig:cor}). The persistence of the shock from the low corona to heliospheric distances indicates that the CME remained strongly super-Alfvénic, driving a large-scale disturbance through the ambient solar wind that was subsequently detected at 1~au.

\subsection{Heliospheric propagation} \label{IS}

\begin{figure*}
    \centering
    \includegraphics[width=0.95\linewidth]{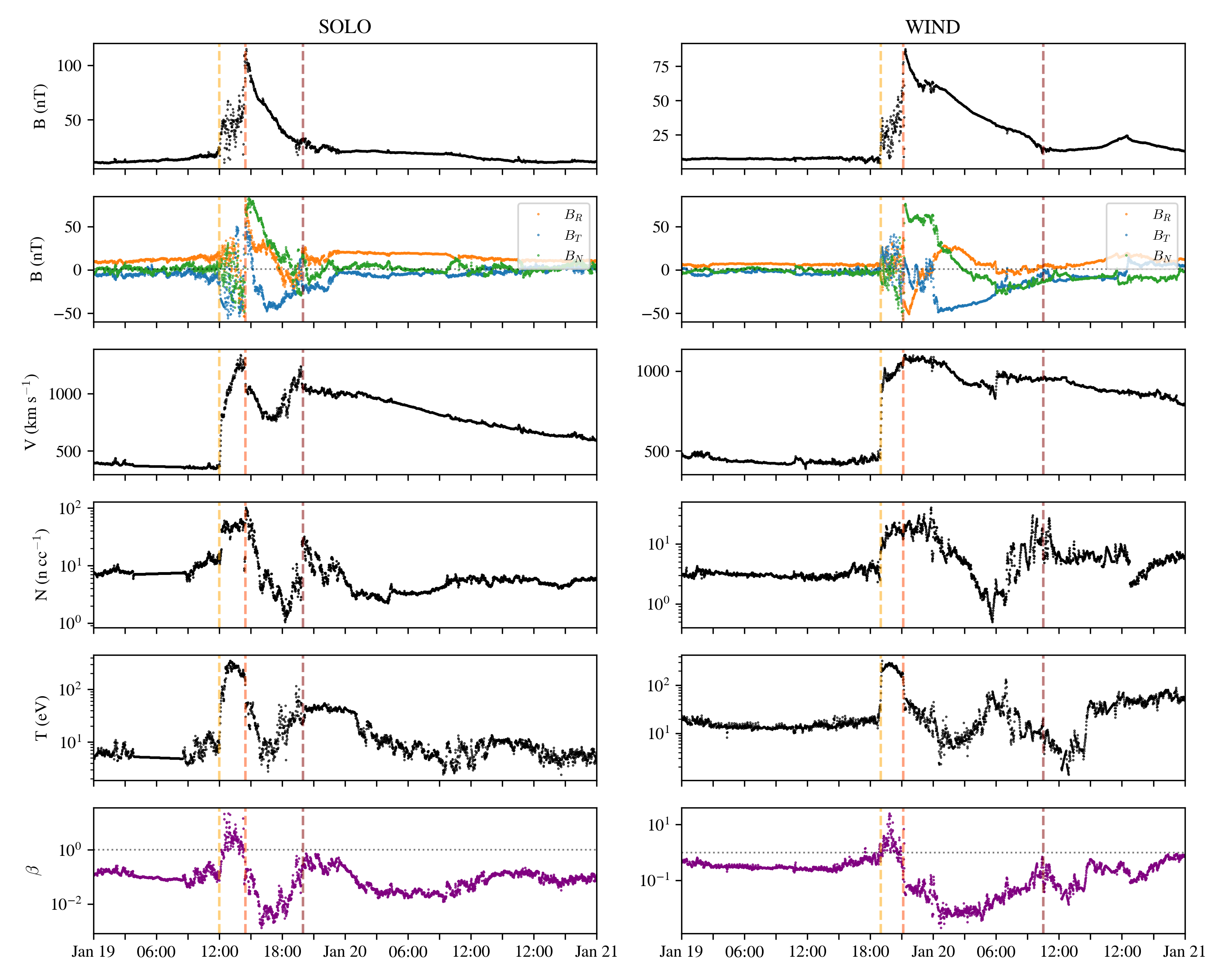}
    \caption{An overivew of the \textit{in situ} observations for the MLK event from Solar Orbiter \textbf{(left)} at $0.75\,$au and $8^\circ$ east of earth, and the Wind spacecraft \textbf{(right)} located at the L1 point. The panels show, in their respective order, the total magnetic field strength $|\mathbf{B}|$, the magnetic field components in the RTN coordinate system, the bulk proton speed $V$, the proton density $N$, the proton temperature $T$ and the plasma beta parameter $\beta$.}
    \label{fig:insitu}
\end{figure*}

The interplanetary CME (ICME) was first observed by SolO, located at 0.75~au and approximately $8^\circ$ east of Earth, at 12:00~UT on January 19, and by the Wind \citep{1997Ogilvie} spacecraft near the L1 point seven hours later, at approximately 19:00~UT.
Figure~\ref{fig:insitu} shows the magnetic field and plasma parameters measured by both spacecraft from January 19 to January 21. For SolO, we use data from the Solar Orbiter Magnetometer \citep[MAG;][]{horbury2020solar} and the Solar Wind Analyser \citep[SWA;][]{owen2020solar}. For Wind, we employ measurements from the Magnetic Field Investigation \citep[MFI;][]{lepping1995} and the Solar Wind Experiment \citep[SWE;][]{ogilvie1995}.
In each panel, the yellow dashed lines mark the arrival time of the sheath, and the orange and maroon dashed lines indicate the start and approximate end times of the magnetic obstacle (MO). The MO is primarily identified using the plasma beta parameter $\beta$, which is defined as the ratio between the plasma and magnetic pressure terms. The preceding solar wind has a speed of $\approx 450\,\textrm{km}\textrm{s}^{-1}$, which abruptly increases to slightly above $1000\,\textrm{km}\textrm{s}^{-1}$ as the ICME shock sweeps over both SolO and the Wind spacecraft. A high-speed stream follows in the wake of the ICME at both locations, with a highly elevated speed that only slowly decays.

At SolO, the sheath region, which follows the shock, lasts about $2\,\textrm{h}$ and $30\,\textrm{m}$ and is followed by a complex MO structure. 
At Wind, the sheath is characterized by an unusually weak magnetic field, and lasts for about $2\,\textrm{h}$ and $13\,\textrm{m}$. Counterintuitively, this is shorter than the duration of the sheath at Solar Orbiter, which is closer to the Sun. The subsequent structure features two distinct regions that are hard to clearly distinguish. Nonetheless, there are strong indications that two separate structures are present, where we denote the first as an intermediate structure and the latter as the primary MO.

The intermediate structure exhibits several characteristics of an MO, including enhanced magnetic field strength and reduced plasma $\beta$, yet also displays unusually high proton density, partially even exceeding that of the sheath, and little evidence of expansion, as indicated by the nearly constant solar wind speed. Beginning near 00:00~UT on January 20, the proton density decreases while the magnetic field components undergo a simultaneous change in their overall trend. The primary MO structure exhibits clearer signatures of an MFR, with an even lower plasma beta, lower proton density, a smoother magnetic field, and some expansion evident from the slope of the speed profile.
The trailing portion of the MO appears to be interacting with a very fast solar wind belonging to a high-speed stream, producing a mixed structure that persists well after the ICME passage.

The Wind observations are consistent with the arrival of two merged magnetic structures, most likely associated to the filament structure identified in the remote sensing images (see Section~\ref{SR}) being compressed and overtaken by the MFR structure related to the main eruption. There is also some small evidence that the intermediate structure may also be present in the SolO observations, but its signatures are much less clear at this location. As such, the SolO observations are more closely represented by the second part of what was observed by Wind, since Solo probably encountered less of the filament structure but still shows some merging signatures.

\section{Modeling}\label{mod}

\begin{figure}
    \centering
    \includegraphics[width=0.95\linewidth]{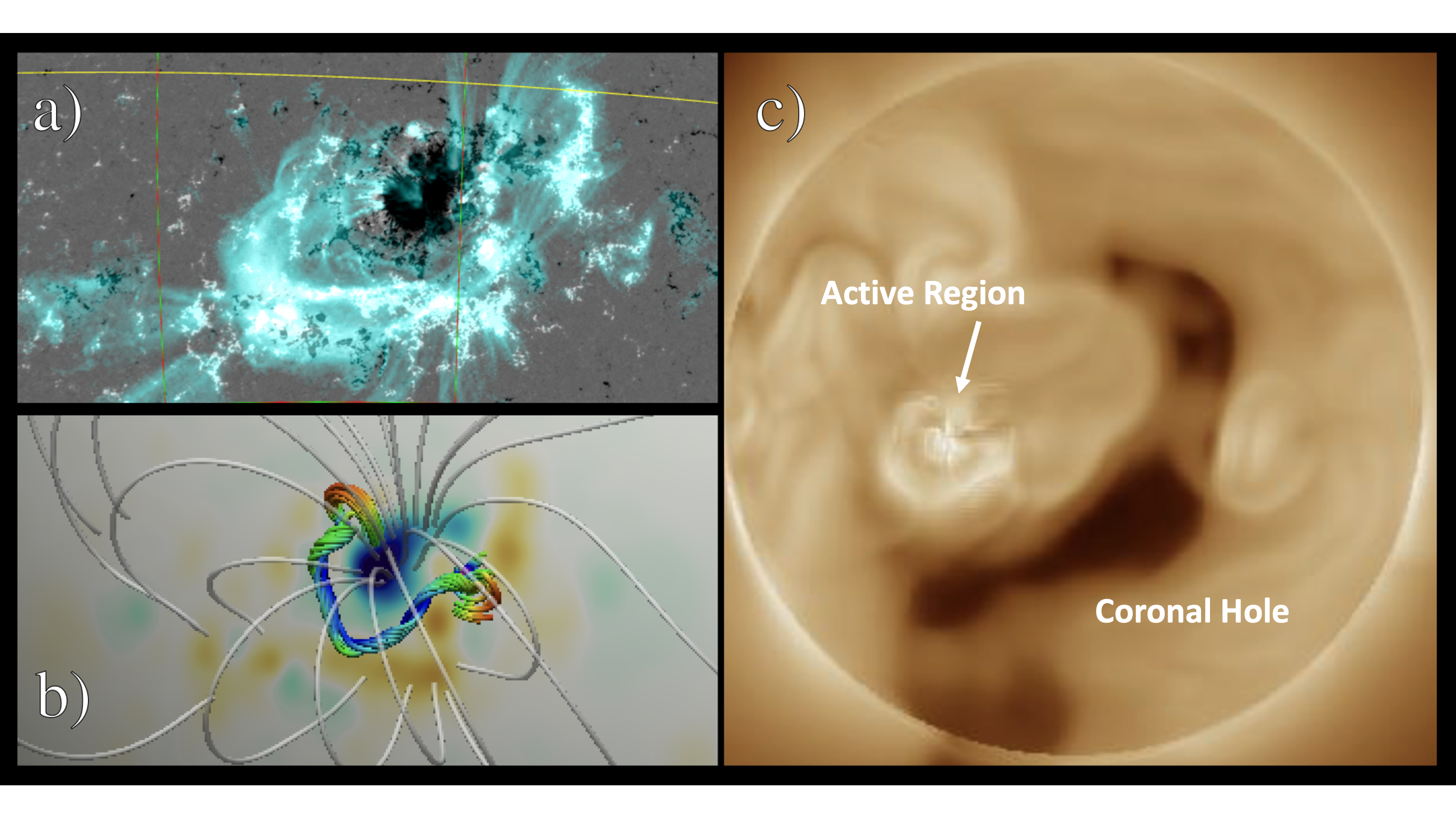}
    \caption{CORHEL-CME input information and initial state. a) Composite AIA 131 and HMI of the hot MFR and source region of the MLK event. b) CORHEL-CME initial MFR field lines (colored field lines) and ambient arcade field lines (white field lines) and surface magnetogram of the source region. c) CORHEL-CME synthetic observable of AIA 195, showcasing the coronal hole and launch site of the eruption (white arrow).}
    \label{fig:ini_model}
\end{figure}

CORHEL-CME\footnote{\url{https://ccmc.gsfc.nasa.gov/models/CORHEL-CME~1/}} \citep{linker_2024} uses the Regularized Biot-Savart Law \citep[RBSL;][]{titov_2018} flux rope model to construct stable MFRs following arbitrary PILs. Once the source region and MFR properties are specified, the model evolves the eruption using thermodynamic MHD from the low corona to beyond 1~au within a self-consistent solar wind solution.

This modeling framework is flexible enough to capture the complexity of CMEs arising from both their internal magnetic structure and their interaction with the ambient coronal field, as demonstrated by \citet{maccormack_2025b, sahade_2026}. Comparisons with observations make it possible to validate both the modeled initial MFR and the surrounding magnetic environment while simultaneously revealing the physical processes governing the CME evolution throughout the inner heliosphere. CORHEL-CME provides a self-consistent simulation of the background magnetic environment derived from the input magnetogram using a coronal heating model. Therefore, in the heliospheric regime (from 30~R$_\odot$ onward), it is not possible to fine-tune the upstream solar wind conditions to match the in situ observations prior to the CME arrival. Instead, the solar wind properties are those produced self-consistently by the thermodynamic coronal MHD model. 

For the base magnetogram, we used a combination of HMI synoptic maps (daily synchronic for January 18, CR2306, and CR2307; the specific map is available with full simulation results) to obtain the most up-to-date representation of the photospheric magnetic field. As mentioned in Section~\ref{SR}, the source region of the CME consists of a hot flux rope (see Fig.~\ref{fig:sr}d) together with quiescent filaments in the surrounding area. We modeled only the hot flux rope, using the observed apex and footpoint locations as constraints. The quiescent filaments were not included because they are rooted in quiet-Sun regions with weak magnetic field strengths and lie along an extended high PIL (about $1.1\,R_\odot$ according to tie-pointing reconstructions). Figure~\ref{fig:ini_model}(a) shows the initial activation of the MFR observed in AIA 131 overlaid on the HMI magnetogram, which was used to determine the PIL and footpoint locations for the CORHEL-CME model shown in Figure~\ref{fig:ini_model}(b). Figure~\ref{fig:ini_model}(c) presents the synthetic AIA 195 emission, illustrating the coronal hole, the global corona, and the activation of the modeled MFR (see Fig.~\ref{fig:sr}(a) for comparison). The complete simulation parameters and results are publicly available at the Community Coordinated Modeling Center's website: \href{https://ccmc.gsfc.nasa.gov/ror/results/viewrun.php?runnumber=Abril_Sahade_040226_SH_1}{MLK}. 

In the next section, we present the modeling results together with a comparison to the remote-sensing and in situ observations. Based on the agreements and discrepancies between the model and the observations, we provide a more complete picture of the evolution of the January 18, 2026 event and the geomagnetic disturbances it produced.

\section{Results}\label{res}

\subsection{Source region and coronal evolution}
\begin{figure*}
    \centering
    \includegraphics[width=0.75\linewidth]{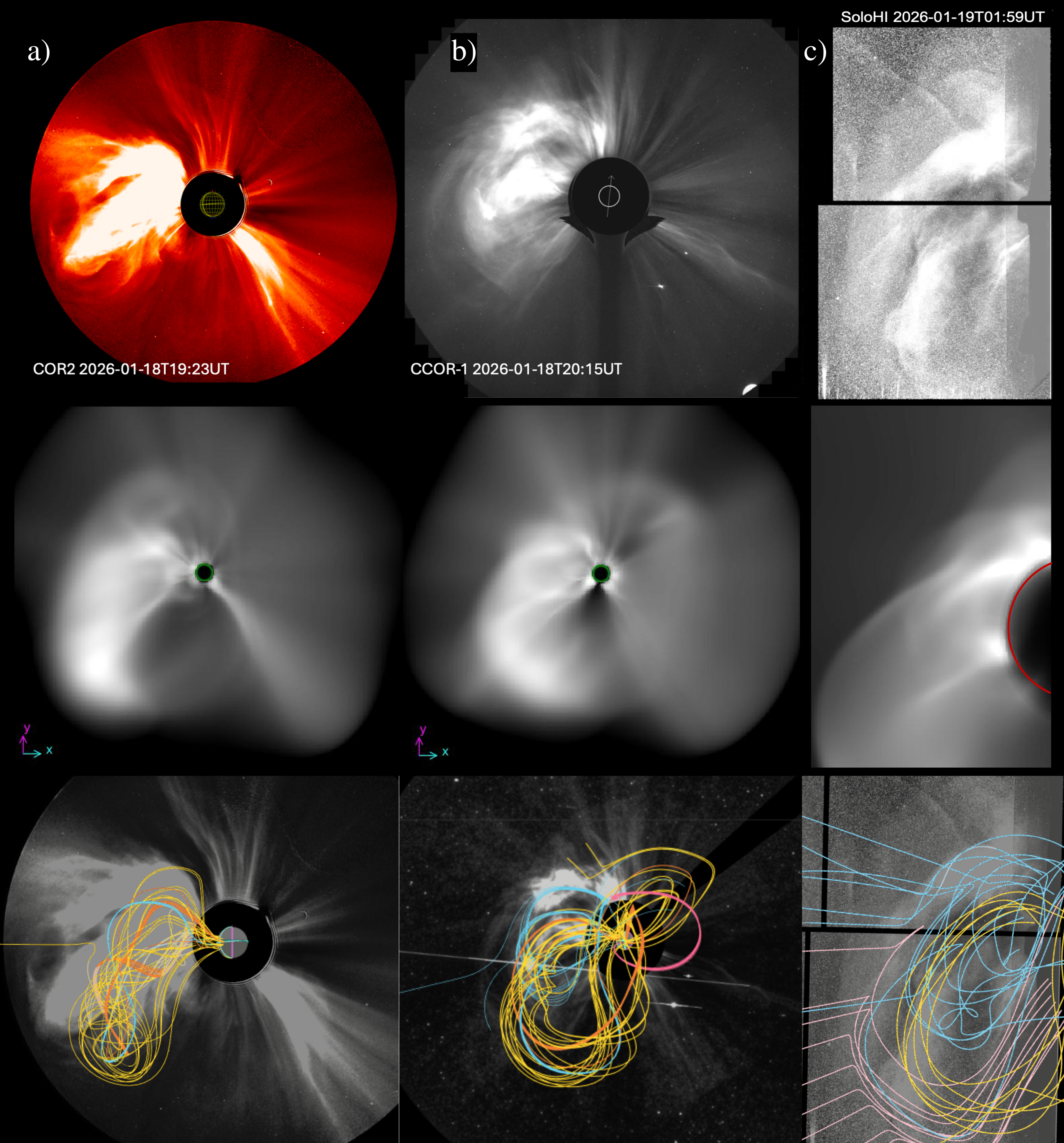}
    \caption{Comparison between the observed and modeled CME morphology from multiple viewpoints. The columns show observations from STA COR2 (a), CCOR-1 (b), and SoloHI tiles 1 and 2 (c). The first row shows the observed white-light images, the second row the corresponding synthetic white-light images, and the third row the modeled magnetic field lines projected onto the observed images. The green circle in the second row represents the $1\,R_\odot$ inner boundary of the coronal simulation, and the red circle on panel (c) represents the $28\,R_\odot$ inner boundary of the heliospheric domain. In the third row, the golden field lines trace the MFR near its axis, while the cyan and pink field lines represent more external magnetic field on the eastern and western legs of the MFR, respectively. The projected magnetic field lines further identify the magnetic structures responsible for the observed density enhancements and provide a three-dimensional interpretation of the CME morphology from the different viewing geometries.}
    \label{fig:cor_model}
\end{figure*}

We first compare the modeled coronal evolution with the multi-viewpoint observations presented in Section~\ref{SR}. By comparing the white-light observations from multiple vantage points with the corresponding synthetic images and the three-dimensional evolution of the modeled MFR, we find overall excellent agreement between the simulation and the observations, as can be noted in Figure~\ref{fig:cor_model}. Figure~\ref{fig:cor_model} shows the CME as observed by STA COR2 (panel a), CCOR-1 (panel b), and SoloHI tiles 1 and 2 (panel c). The first row shows the white-light images, the second row the corresponding synthetic images, and the third row the magnetic field lines projected onto the white-light images to facilitate comparison between the observed density enhancements and the underlying magnetic structure. The golden lines correspond to the MFR field near its axis, cyan field lines correspond to more external field on the eastern leg of the MFR, and pink lines correspond to external field on the western leg of the MFR. The model reproduces the global morphology of the CME, including its asymmetric expansion, the flattened southern flank produced by the nearby coronal hole, and several internal density structures visible from different viewing geometries. The only major observational feature absent from the synthetic images is the northern overdense component identified in the coronagraph observations, which we interpret as filament material. This discrepancy is expected because the quiescent, high-density filaments surrounding the hot MFR were not included in the initial magnetic configuration of the simulation.

The agreement between the observations and the model provides confidence in the reconstructed three-dimensional geometry of the eruption and motivates a closer examination of the magnetic environment that controls its evolution. The erupting MFR that produced the MLK event originated from a strong circular AR, with its central spine rooted in its negative polarity (see Fig.~\ref{fig:ini_model}). To the west and south, a positive-polarity open field associated with the coronal hole surrounded the pseudostreamer-like topology, while closed arcades belonging to a helmet streamer were present to the northeast. The U-shaped, right-handed MFR evolved within this asymmetric magnetic environment. As the closed magnetic field of the pseudostreamer-like topology reconnected beneath the rising MFR, the overlying arcades added magnetic flux to the MFR while simultaneously forming post-flare arcades on the opposite side. The western and southern segments of the MFR rose and expanded more rapidly because of the weaker confinement provided by the overlying open field. This behavior is consistent with the observed west-to-east progression of the flare ribbons and with the greater leading height of the CME front toward the south, as observed in coronagraph images from the STEREO-A viewpoint (Figure~\ref{fig:cor}b).

Although the initial shape of the MFR favored southward evolution, the open magnetic flux of the coronal hole constrained and redirected its western and southern expansion. The favorable magnetic polarity alignment reduced reconnection between the MFR and the open field, thereby minimizing MFR erosion. As a result, the MFR underwent rapid expansion with only minimal deceleration in the coronal domain.

The synthetic CME (middle row of Figure~\ref{fig:cor_model}) exhibits slightly greater southward expansion than observed, as also evident from the difference in the projected field lines in the third row of Figure~\ref{fig:cor_model}a. This discrepancy is most likely associated with an underestimation of the magnetic strength of the polar segment of the coronal hole in the reconstructed background corona, resulting in weaker confinement of the southern flank of the eruption. This interpretation is further supported by the comparison between the observed EUV corona (Figure~\ref{fig:sr}d) and the synthetic AIA 195~\AA{} emission (Figure~\ref{fig:ini_model}c), in which the simulated coronal hole appears smaller than observed.

The three-dimensional magnetic field reconstruction also provides a direct interpretation of the white-light morphology (third row of Figure~\ref{fig:cor_model}). By projecting the modeled MFR field lines onto the coronagraph images, we identify the origin of the different components of the observed CME. From the Earth and Solar Orbiter viewpoints, the brightest emission is primarily associated with the eastern segment of the MFR, whereas the western segment propagates largely toward Earth and therefore contributes comparatively little brightness because of Thomson-scattering effects. The northern portion of the CME-driven shock corresponds to the projected western segment of the MFR and its interaction with the western boundary of the coronal hole, producing the sharp-edged morphology observed along the northern shock front.
In particular, the comparison indicates that the western leg of the MFR, together with the adjacent coronal-hole magnetic flux, are directed toward Earth. This initial magnetic configuration provides the basis for interpreting the CME’s subsequent evolution and its magnetic orientation at 1~au.

\subsection{Heliospheric propagation and impact}
\begin{figure*}
    \centering
    \includegraphics[width=0.95\linewidth]{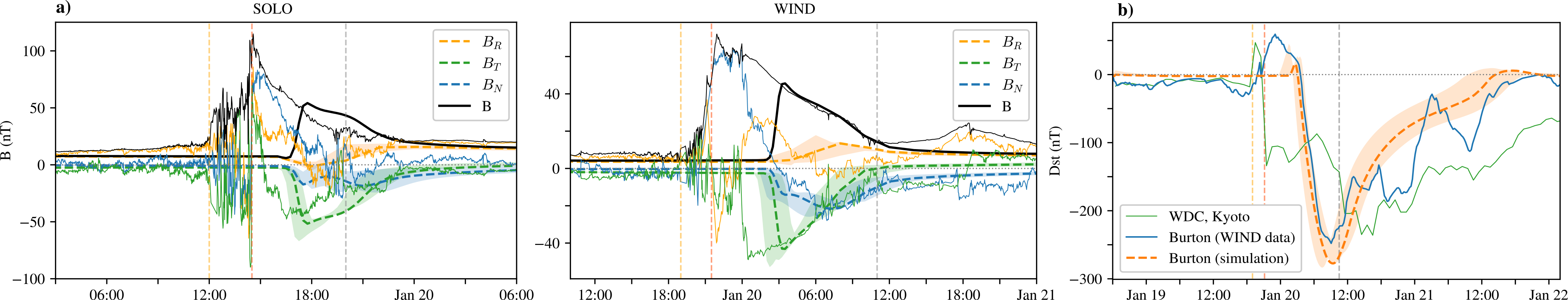}
    \caption{Comparison between the simulated and observed heliospheric impact of the MLK event. (a) In situ magnetic field profiles measured by SolO and Wind (thin solid lines) and corresponding synthetic profiles from the simulation (dashed lines), with the simulated magnetic field strength scaled by a factor of 2. Shaded regions indicate the uncertainty associated with a $\pm5^\circ$ variation in the synthetic spacecraft position. Vertical lines mark the observed boundaries used for comparison. (b) Observed Dst index from the WDC for Geomagnetism, Kyoto (solid green), Dst calculated from the Wind magnetic field and plasma measurements using the \citet{burton_1975} formulation (solid blue), and Dst derived from the simulated solar wind parameters (dashed orange), with the orange shading indicating the uncertainty associated with the synthetic spacecraft position. The vertical lines correspond to those in panel (a) for Wind.}
    \label{fig:insitu_model}
\end{figure*}

The simulated CME maintains its coherent magnetic structure throughout its propagation to 1~au and does not undergo significant deformation or compression by interaction with stream interaction regions or other large-scale solar wind structures. 

We next compare the simulated heliospheric evolution with the magnetic in situ observations described in Section~\ref{IS}.
Figure~\ref{fig:insitu_model}a presents the in situ magnetic field profiles measured by both SolO and Wind. The observations are shown as thin solid lines, while the corresponding synthetic magnetic profiles at each spacecraft location are shown as dashed lines. The shaded regions indicate the uncertainty associated with a $\pm5^\circ$ variation in the synthetic spacecraft position. The simulated magnetic field strength is enhanced by a factor of 2. Such an underestimation of the magnetic field strength at 1~au is a known limitation of the CORHEL-CME framework, with previous studies reporting discrepancies of a factor of $\sim$1.5--2 \citep{linker_2016,torok_2018,sahade_2026}. This discrepancy is likely related to model simplifications and, in particular, to an underestimation of the photospheric magnetic field strength in the input magnetograms. A factor of 2 provides consistent agreement with the observations at both spacecraft locations (0.74 au and 1 au), supporting the robustness of this scaling choice. 
Although the nominal model output has a temporal resolution of 2 hours, we advected the solution using the finer spatial resolution of the simulation, particularly in the radial direction, to reconstruct synthetic in situ profiles at a cadence of 15 minutes. Overall, the simulated magnetic profiles reproduce the second half of the MO remarkably well at both spacecraft. Moreover, despite sampling additional locations throughout the 3D simulation domain, we did not identify any position that provided a better match to the observed magnetic profiles than the nominal spacecraft trajectories.

As discussed in Section~\ref{IS}, the in situ measurements indicate the arrival of a complex, and most likely mixed, magnetic structure. We proposed that this complexity resulted from the eruption of a long filament located along the western leg of the erupting hot flux rope. This component is absent from the simulation because the filament eruption was not included in the model. Nevertheless, the good agreement between the simulated and observed magnetic profiles during the latter portion of the MO provides support for the simulated flux-rope interpretation. We therefore computed the Dst index using the empirical formulation of \citet{burton_1975} to further investigate the origin of the severe geomagnetic response associated with the MLK event.

Figure~\ref{fig:insitu_model}b compares the observed Dst index from the World Data Center (WDC) for Geomagnetism, Kyoto (solid green line), the Dst index calculated using the Burton model driven by the Wind observations at 15 minutes cadence (solid blue line), and the Dst index calculated from the simulated plasma and magnetic field parameters (dashed orange line), together with the associated uncertainty (orange shaded region). The vertical lines are the same as in panel (a) for Wind data to guide the reader in the time comparison since timescales differ. The observed Dst profile (green line) exhibits a characteristic two-step decrease (about Jan 19 at 19:00~UT and 21:15~UT) that is not reproduced by the Burton model driven by the L1 observations, suggesting that this feature may result from magnetospheric processes not represented by the empirical formulation. Nevertheless, the Dst index computed from the Wind measurements (blue line) successfully reproduces the main decrease to approximately $-200,\mathrm{nT}$ (near the grey vertical line), as well as most of the subsequent gradual recovery. Likewise, the Dst index derived from the simulation (orange line) closely follows both the main depression and the recovery phase, showing excellent agreement with the Dst calculated from the Wind observations (see blue and orange curves).
As expected, the main decrease in the Dst index is associated with the sustained southward magnetic field carried by the main MFR, which is well reproduced by the simulation. The subsequent gradual recovery is associated with the wake of the ICME and the arrival of the trailing high-speed stream originating from the adjacent coronal hole.
The close agreement between the simulated and observed Dst evolution indicates that the main features of the storm intensity and recovery are captured by the simulated conditions. The physical interpretation of these results and respective roles of the CME magnetic configuration and the trailing high-speed stream in controlling the intensity and duration of the storm are further discussed in next section.

\section{Discussion and conclusions}\label{conc}
In this work, we presented a comprehensive analysis of the 18--19 January 2026 (MLK) space weather event by combining remote-sensing observations, in situ measurements, and physics-based simulations with CORHEL-CME. This integrated approach enabled us to reconstruct the evolution of the eruption continuously from its initiation in the low corona to its impact at Earth and to identify the physical mechanisms responsible for one of the most intense geomagnetic storms of Solar Cycle 25.

The observations reveal that the eruption originated from a right-handed hot MFR embedded within a complex magnetic environment composed of a pseudostreamer-like closed structure and surrounding coronal hole. Prior to the eruption, the active region underwent a sequence of confined and eruptive flares that progressively destabilized the magnetic configuration. The subsequent eruption interacted with nearby quiescent filaments, producing a complex CME morphology that was later reflected in the in situ observations as a mixed magnetic structure. The absence of this filament material in the numerical simulation naturally explains the primary discrepancy between the synthetic and observed white-light images, as well as the differences in the leading portion of the magnetic obstacle measured at 1~au.
The combination of multi-viewpoint observations and three-dimensional MHD modeling provides a coherent interpretation of the coronal evolution of the event. The simulation successfully reproduces the global morphology of the CME, including its asymmetric expansion, the flattened southern flank, the magnetic field orientation, and the magnetic profiles measured by both SolO and Wind. The overall agreement between the synthetic observables and the observations from multiple independent instruments gives confidence that the reconstructed magnetic configuration accurately represents the large-scale evolution of the eruption and supports the physical interpretation presented throughout this work.

One of the most important results concerns the role of the adjacent coronal hole in controlling the evolution of the CME. The simulation indicates that the expanding MFR encountered open magnetic flux with the same magnetic polarity as its western leg. This favorable magnetic alignment substantially reduced magnetic erosion through interchange reconnection, allowing the CME to preserve its magnetic integrity during its escape from the low corona. At the same time, the open-field configuration provided relatively weak magnetic confinement, permitting rapid expansion with only limited deceleration in the coronal domain. These effects allowed the eruption to maintain both a strong magnetic field and a high propagation speed before entering the heliosphere, preconditioning the ICME for an intense geomagnetic impact.

The heliospheric simulations further show that, after leaving the corona, the CME propagated without undergoing major interactions with other large-scale solar wind structures capable of significantly modifying its magnetic configuration. Consequently, the strong southward magnetic field responsible for the geomagnetic storm was largely inherited from the coronal evolution of the eruption rather than being produced by processes occurring during interplanetary propagation. The close agreement between the simulated magnetic profiles and those measured in situ, together with the successful reproduction of the Dst evolution using the simulated plasma and magnetic field parameters, strongly supports this interpretation.
Although the peak intensity of the geomagnetic storm was primarily controlled by the magnetic configuration of the MFR, the duration of the disturbance appears to have been influenced by the ambient solar wind environment. Both the observations and the simulations indicate that the trailing portion of the magnetic obstacle merged with the leading edge of a high-speed stream originating from the same coronal hole that influenced the early CME evolution. This interaction likely prolonged the interval of enhanced solar wind driving, delaying the recovery of the magnetosphere and contributing to the extended duration of the G4 geomagnetic storm. 
The MLK event therefore illustrates that the intensity and duration of severe geomagnetic storms may be governed by different physical processes: the former by the magnetic configuration established during the eruption, and the latter by the large-scale solar wind environment encountered during propagation.

Beyond characterizing this individual event, this study demonstrates the value of combining physics-based modeling with coordinated remote-sensing and in situ observations to establish a physically consistent Sun-to-Earth interpretation of extreme space weather events. 
The present work focused primarily on understanding the origin of the geomagnetic disturbance associated with the MLK event. However, the eruption also produced one of the most significant solar radiation storms of Solar Cycle 25. A detailed investigation of the particle acceleration processes responsible for the observed SEP event, together with the characterization of the associated CME-driven shock and its evolution throughout the inner heliosphere, will be presented in a forthcoming study.

%
%

%
%

\begin{acknowledgments}


\end{acknowledgments}
A.S.\ is supported by an appointment to the NASA Postdoctoral Program at the NASA Goddard Space Flight Center, administered by Oak Ridge Associated Universities under contract with NASA.
C.M.\ is supported by the NASA Heliophysics Division, Solar Orbiter Collaboration Office under DPR NNG09EK11I. 
A.J.W.\, A.N.\, D.G.L., and T.N.-C.\ thank the support of the Solar Orbiter mission. 
A.J.W.\ and E.P.\ acknowledge support from NASA's HSR (Grant 80NSSC25M7101) program.
E.P.\ also acknowledges support from NASA's LWS (Grant 80NSSC24K1108) and LWS-SC (Grant 80NSSC22K0893) programs; support was also provided by NASA under Grant 80NSSC26K0467, administered by the NASA Goddard Space Flight Center as part of the Solar Orbiter project. 
Simulation results have been provided by the Community Coordinated Modeling Center (CCMC) at Goddard Space Flight Center through their publicly available simulation services (https://ccmc.gsfc.nasa.gov). The MHDweb tool \citep{davidson2025} was used to assist with the processing of synthetic observables (https://predsci.com/mhdweb2).
Solar Orbiter is a mission of international cooperation between the European Space Agency (ESA) and the National Aeronautics and Space Administration (NASA), operated by ESA. The Solar Orbiter Heliospheric Imager (SoloHI) instrument was designed, built, and is now operated by the U.S.\ Naval Research Laboratory with the support of the NASA Heliophysics Division, Solar Orbiter Collaboration Office under DPR NNG09EK11I.

\begin{contribution}

A.S.\ conceived and led the study, performed the CORHEL–CME simulations and associated analyses, and conducted the comparison with observational data. C.M.\ contributed to the development of the study and the analysis of remote-sensing observations. A.J.W.\ contributed to the analysis and interpretation of the in situ observations. A.N.\ and D.G.L.\ contributed to the scientific interpretation and development of the research concept. E.P.\ provided the merged magnetogram used for the CORHEL–CME simulations and contributed to the analysis and manuscript preparation. T.N.C.\ contributed to manuscript preparation. All authors contributed to the interpretation and discussion of the results and to the revision of the manuscript.

\end{contribution}

\facilities{SDO/AIA, GOES/CCOR-1, STEREO/COR2, SolO/SoloHI, Solo/MAG, Solo/SWA, Wind/MFI, and Wind/SWE data}

%
%

\bibliography{bio}

\begin{thebibliography}{}
\expandafter\ifx\csname natexlab\endcsname\relax\def\natexlab#1{#1}\fi
\providecommand{\url}[1]{\href{#1}{#1}}
\providecommand{\dodoi}[1]{doi:~\href{http://doi.org/#1}{\nolinkurl{#1}}}
\providecommand{\doeprint}[1]{\href{http://ascl.net/#1}{\nolinkurl{http://ascl.net/#1}}}
\providecommand{\doarXiv}[1]{\href{https://arxiv.org/abs/#1}{\nolinkurl{https://arxiv.org/abs/#1}}}

\bibitem[{R.~K. {Burton} {et~al.}(1975){Burton}, {McPherron}, \& {Russell}}]{burton_1975}
{Burton}, R.~K., {McPherron}, R.~L., \& {Russell}, C.~T. 1975, \bibinfo{title}{{An empirical relationship between interplanetary conditions and Dst},} \jgr, 80, 4204, \dodoi{10.1029/JA080i031p04204}

\bibitem[{P.~F. {Chen}(2011){Chen}}]{chen_2011}
{Chen}, P.~F. 2011, \bibinfo{title}{{Coronal Mass Ejections: Models and Their Observational Basis},} Living Reviews in Solar Physics, 8, 1, \dodoi{10.12942/lrsp-2011-1}

\bibitem[{I.~A. {Daglis} {et~al.}(1999){Daglis}, {Thorne}, {Baumjohann}, \& {Orsini}}]{1999Daglis}
{Daglis}, I.~A., {Thorne}, R.~M., {Baumjohann}, W., \& {Orsini}, S. 1999, \bibinfo{title}{{The terrestrial ring current: Origin, formation, and decay},} Reviews of Geophysics, 37, 407, \dodoi{10.1029/1999RG900009}

\bibitem[{R. {Davidson} {et~al.}(2025){Davidson}, {Reyes}, {Downs}, {Linker}, {Riley}, {Caplan}, {Ben-Nun}, \& {Palmerio}}]{davidson2025}
{Davidson}, R., {Reyes}, A., {Downs}, C., {et~al.} 2025, in AGU Fall Meeting Abstracts, Vol. 2025, AGU Fall Meeting Abstracts, SA31D--2342

\bibitem[{J.~P. {Eastwood} {et~al.}(2017){Eastwood}, {Biffis}, {Hapgood}, {Green}, {Bisi}, {Bentley}, {Wicks}, {McKinnell}, {Gibbs}, \& {Burnett}}]{eastwood_2017}
{Eastwood}, J.~P., {Biffis}, E., {Hapgood}, M.~A., {et~al.} 2017, \bibinfo{title}{{The Economic Impact of Space Weather: Where Do We Stand?},} RiskA, 37, 206, \dodoi{10.1111/risa.12765}

\bibitem[{L. {Fletcher} {et~al.}(2011){Fletcher}, {Dennis}, {Hudson}, {Krucker}, {Phillips}, {Veronig}, {Battaglia}, {Bone}, {Caspi}, {Chen}, {Gallagher}, {Grigis}, {Ji}, {Liu}, {Milligan}, \& {Temmer}}]{fletcher_2011}
{Fletcher}, L., {Dennis}, B.~R., {Hudson}, H.~S., {et~al.} 2011, \bibinfo{title}{{An Observational Overview of Solar Flares},} \ssr, 159, 19, \dodoi{10.1007/s11214-010-9701-8}

\bibitem[{W.~D. {Gonzalez} {et~al.}(1999){Gonzalez}, {Tsurutani}, \& {Cl{\'u}a de Gonzalez}}]{1999Gonzalez}
{Gonzalez}, W.~D., {Tsurutani}, B.~T., \& {Cl{\'u}a de Gonzalez}, A.~L. 1999, \bibinfo{title}{{Interplanetary origin of geomagnetic storms},} \ssr, 88, 529, \dodoi{10.1023/A:1005160129098}

\bibitem[{N. {Gopalswamy} {et~al.}(2005){Gopalswamy}, {Yashiro}, {Liu}, {Michalek}, {Vourlidas}, {Kaiser}, \& {Howard}}]{gopalswamy_2005}
{Gopalswamy}, N., {Yashiro}, S., {Liu}, Y., {et~al.} 2005, \bibinfo{title}{{Coronal mass ejections and other extreme characteristics of the 2003 October-November solar eruptions},} Journal of Geophysical Research (Space Physics), 110, A09S15, \dodoi{10.1029/2004JA010958}

\bibitem[{T.~S. {Horbury} {et~al.}(2020){Horbury}, {O'Brien}, {Carrasco Blazquez}, {Bendyk}, {Brown}, {Hudson}, {Evans}, {Oddy}, {Carr}, {Beek}, {Cupido}, {Bhattacharya}, {Dominguez}, {Matthews}, {Myklebust}, {Whiteside}, {Bale}, {Baumjohann}, {Burgess}, {Carbone}, {Cargill}, {Eastwood}, {Erd{\"o}s}, {Fletcher}, {Forsyth}, {Giacalone}, {Glassmeier}, {Goldstein}, {Hoeksema}, {Lockwood}, {Magnes}, {Maksimovic}, {Marsch}, {Matthaeus}, {Murphy}, {Nakariakov}, {Owen}, {Owens}, {Rodriguez-Pacheco}, {Richter}, {Riley}, {Russell}, {Schwartz}, {Vainio}, {Velli}, {Vennerstrom}, {Walsh}, {Wimmer-Schweingruber}, {Zank}, {M{\"u}ller}, {Zouganelis}, \& {Walsh}}]{horbury2020solar}
{Horbury}, T.~S., {O'Brien}, H., {Carrasco Blazquez}, I., {et~al.} 2020, \bibinfo{title}{{The Solar Orbiter magnetometer},} \aap, 642, A9, \dodoi{10.1051/0004-6361/201937257}

\bibitem[{R.~A. {Howard} {et~al.}(2008){Howard}, {Moses}, {Vourlidas}, {Newmark}, {Socker}, {Plunkett}, {Korendyke}, {Cook}, {Hurley}, {Davila}, {Thompson}, {St Cyr}, {Mentzell}, {Mehalick}, {Lemen}, {Wuelser}, {Duncan}, {Tarbell}, {Wolfson}, {Moore}, {Harrison}, {Waltham}, {Lang}, {Davis}, {Eyles}, {Mapson-Menard}, {Simnett}, {Halain}, {Defise}, {Mazy}, {Rochus}, {Mercier}, {Ravet}, {Delmotte}, {Auchere}, {Delaboudiniere}, {Bothmer}, {Deutsch}, {Wang}, {Rich}, {Cooper}, {Stephens}, {Maahs}, {Baugh}, {McMullin}, \& {Carter}}]{howard2008sun}
{Howard}, R.~A., {Moses}, J.~D., {Vourlidas}, A., {et~al.} 2008, \bibinfo{title}{{Sun Earth Connection Coronal and Heliospheric Investigation (SECCHI)},} \ssr, 136, 67, \dodoi{10.1007/s11214-008-9341-4}

\bibitem[{R.~A. {Howard} {et~al.}(2020){Howard}, {Vourlidas}, {Colaninno}, {Korendyke}, {Plunkett}, {Carter}, {Wang}, {Rich}, {Lynch}, {Thurn}, {Socker}, {Thernisien}, {Chua}, {Linton}, {Koss}, {Tun-Beltran}, {Dennison}, {Stenborg}, {McMullin}, {Hunt}, {Baugh}, {Clifford}, {Keller}, {Janesick}, {Tower}, {Grygon}, {Farkas}, {Hagood}, {Eisenhauer}, {Uhl}, {Yerushalmi}, {Smith}, {Liewer}, {Velli}, {Linker}, {Bothmer}, {Rochus}, {Halain}, {Lamy}, {Auch{\`e}re}, {Harrison}, {Rouillard}, {Patsourakos}, {St. Cyr}, {Gilbert}, {Maldonado}, {Mariano}, \& {Cerullo}}]{howard_2020}
{Howard}, R.~A., {Vourlidas}, A., {Colaninno}, R.~C., {et~al.} 2020, \bibinfo{title}{{The Solar Orbiter Heliospheric Imager (SoloHI)},} \aap, 642, A13, \dodoi{10.1051/0004-6361/201935202}

\bibitem[{A.~W. {James} {et~al.}(2020){James}, {Green}, {van Driel-Gesztelyi}, \& {Valori}}]{2020James}
{James}, A.~W., {Green}, L.~M., {van Driel-Gesztelyi}, L., \& {Valori}, G. 2020, \bibinfo{title}{{A new trigger mechanism for coronal mass ejections. The role of confined flares and photospheric motions in the formation of hot flux ropes},} \aap, 644, A137, \dodoi{10.1051/0004-6361/202038781}

\bibitem[{M.~L. {Kaiser} {et~al.}(2008){Kaiser}, {Kucera}, {Davila}, {St. Cyr}, {Guhathakurta}, \& {Christian}}]{kaiser2008stereo}
{Kaiser}, M.~L., {Kucera}, T.~A., {Davila}, J.~M., {et~al.} 2008, \bibinfo{title}{{The STEREO Mission: An Introduction},} \ssr, 136, 5, \dodoi{10.1007/s11214-007-9277-0}

\bibitem[{Y. {Kamide} {et~al.}(1998){Kamide}, {Baumjohann}, {Daglis}, {Gonzalez}, {Grande}, {Joselyn}, {McPherron}, {Phillips}, {Reeves}, {Rostoker}, {Sharma}, {Singer}, {Tsurutani}, \& {Vasyliunas}}]{kamide_1998}
{Kamide}, Y., {Baumjohann}, W., {Daglis}, I.~A., {et~al.} 1998, \bibinfo{title}{{Current understanding of magnetic storms: Storm-substorm relationships},} \jgr, 103, 17705, \dodoi{10.1029/98JA01426}

\bibitem[{B. {Kliem} {et~al.}(2021){Kliem}, {Lee}, {Liu}, {White}, {Liu}, \& {Masuda}}]{kliem_2021}
{Kliem}, B., {Lee}, J., {Liu}, R., {et~al.} 2021, \bibinfo{title}{{Nonequilibrium Flux Rope Formation by Confined Flares Preceding a Solar Coronal Mass Ejection},} \apj, 909, 91, \dodoi{10.3847/1538-4357/abda37}

\bibitem[{J.~R. {Lemen} {et~al.}(2012){Lemen}, {Title}, {Akin}, {Boerner}, {Chou}, {Drake}, {Duncan}, {Edwards}, {Friedlaender}, {Heyman}, {Hurlburt}, {Katz}, {Kushner}, {Levay}, {Lindgren}, {Mathur}, {McFeaters}, {Mitchell}, {Rehse}, {Schrijver}, {Springer}, {Stern}, {Tarbell}, {Wuelser}, {Wolfson}, {Yanari}, {Bookbinder}, {Cheimets}, {Caldwell}, {Deluca}, {Gates}, {Golub}, {Park}, {Podgorski}, {Bush}, {Scherrer}, {Gummin}, {Smith}, {Auker}, {Jerram}, {Pool}, {Soufli}, {Windt}, {Beardsley}, {Clapp}, {Lang}, \& {Waltham}}]{lemen2012atmospheric}
{Lemen}, J.~R., {Title}, A.~M., {Akin}, D.~J., {et~al.} 2012, \bibinfo{title}{{The Atmospheric Imaging Assembly (AIA) on the Solar Dynamics Observatory (SDO)},} \solphys, 275, 17, \dodoi{10.1007/s11207-011-9776-8}

\bibitem[{R.~P. {Lepping} {et~al.}(1995){Lepping}, {Ac{\~{u}}na}, {Burlaga}, {Farrell}, {Slavin}, {Schatten}, {Mariani}, {Ness}, {Neubauer}, {Whang}, {Byrnes}, {Kennon}, {Panetta}, {Scheifele}, \& {Worley}}]{lepping1995}
{Lepping}, R.~P., {Ac{\~{u}}na}, M.~H., {Burlaga}, L.~F., {et~al.} 1995, \bibinfo{title}{{The Wind Magnetic Field Investigation},} \ssr, 71, 207, \dodoi{10.1007/BF00751330}

\bibitem[{J. {Linker} {et~al.}(2016){Linker}, {Torok}, {Downs}, {Lionello}, {Titov}, {Caplan}, {Miki{\'c}}, \& {Riley}}]{linker_2016}
{Linker}, J., {Torok}, T., {Downs}, C., {et~al.} 2016, in American Institute of Physics Conference Series, Vol. 1720, Solar Wind 14 (AIP), 020002, \dodoi{10.1063/1.4943803}

\bibitem[{J.~A. {Linker} {et~al.}(2024){Linker}, {Torok}, {Downs}, {Caplan}, {Titov}, {Reyes}, {Lionello}, \& {Riley}}]{linker_2024}
{Linker}, J.~A., {Torok}, T., {Downs}, C., {et~al.} 2024, in Journal of Physics Conference Series, Vol. 2742, Journal of Physics Conference Series (IOP), 012012, \dodoi{10.1088/1742-6596/2742/1/012012}

\bibitem[{C. {Mac Cormack} {et~al.}(2025){Mac Cormack}, {Sahade}, {Vourlidas}, {Hess}, {Colaninno}, \& {Nieves-Chinchilla}}]{maccormack_2025b}
{Mac Cormack}, C., {Sahade}, A., {Vourlidas}, A., {et~al.} 2025, \bibinfo{title}{{Understanding the Complex Morphology of a Coronal Mass Ejection: Multiview Analysis and Numerical Modeling},} \apj, 988, 180, \dodoi{10.3847/1538-4357/ade311}

\bibitem[{D. {M{\"u}ller} {et~al.}(2020){M{\"u}ller}, {St. Cyr}, {Zouganelis}, {Gilbert}, {Marsden}, {Nieves-Chinchilla}, {Antonucci}, {Auch{\`e}re}, {Berghmans}, {Horbury}, {Howard}, {Krucker}, {Maksimovic}, {Owen}, {Rochus}, {Rodriguez-Pacheco}, {Romoli}, {Solanki}, {Bruno}, {Carlsson}, {Fludra}, {Harra}, {Hassler}, {Livi}, {Louarn}, {Peter}, {Sch{\"u}hle}, {Teriaca}, {del Toro Iniesta}, {Wimmer-Schweingruber}, {Marsch}, {Velli}, {De Groof}, {Walsh}, \& {Williams}}]{muller_2020}
{M{\"u}ller}, D., {St. Cyr}, O.~C., {Zouganelis}, I., {et~al.} 2020, \bibinfo{title}{{The Solar Orbiter mission. Science overview},} \aap, 642, A1, \dodoi{10.1051/0004-6361/202038467}

\bibitem[{A. {Nindos} {et~al.}(2020){Nindos}, {Patsourakos}, {Vourlidas}, {Cheng}, \& {Zhang}}]{nindos_2020}
{Nindos}, A., {Patsourakos}, S., {Vourlidas}, A., {Cheng}, X., \& {Zhang}, J. 2020, \bibinfo{title}{{When do solar erupting hot magnetic flux ropes form?},} \aap, 642, A109, \dodoi{10.1051/0004-6361/202038832}

\bibitem[{K.~W. {Ogilvie} \& M.~D. {Desch}(1997){Ogilvie} \& {Desch}}]{1997Ogilvie}
{Ogilvie}, K.~W., \& {Desch}, M.~D. 1997, \bibinfo{title}{{The wind spacecraft and its early scientific results},} AdSpR, 20, 559, \dodoi{10.1016/S0273-1177(97)00439-0}

\bibitem[{K.~W. {Ogilvie} {et~al.}(1995){Ogilvie}, {Chornay}, {Fritzenreiter}, {Hunsaker}, {Keller}, {Lobell}, {Miller}, {Scudder}, {Sittler}, {Torbert}, {Bodet}, {Needell}, {Lazarus}, {Steinberg}, {Tappan}, {Mavretic}, \& {Gergin}}]{ogilvie1995}
{Ogilvie}, K.~W., {Chornay}, D.~J., {Fritzenreiter}, R.~J., {et~al.} 1995, \bibinfo{title}{{SWE, A Comprehensive Plasma Instrument for the Wind Spacecraft},} \ssr, 71, 55, \dodoi{10.1007/BF00751326}

\bibitem[{C.~J. {Owen} {et~al.}(2020){Owen}, {Bruno}, {Livi}, {Louarn}, {Al Janabi}, {Allegrini}, {Amoros}, {Baruah}, {Barthe}, {Berthomier}, {Bordon}, {Brockley-Blatt}, {Brysbaert}, {Capuano}, {Collier}, {DeMarco}, {Fedorov}, {Ford}, {Fortunato}, {Fratter}, {Galvin}, {Hancock}, {Heirtzler}, {Kataria}, {Kistler}, {Lepri}, {Lewis}, {Loeffler}, {Marty}, {Mathon}, {Mayall}, {Mele}, {Ogasawara}, {Orlandi}, {Pacros}, {Penou}, {Persyn}, {Petiot}, {Phillips}, {P{\v{r}}ech}, {Raines}, {Reden}, {Rouillard}, {Rousseau}, {Rubiella}, {Seran}, {Spencer}, {Thomas}, {Trevino}, {Verscharen}, {Wurz}, {Alapide}, {Amoruso}, {Andr{\'e}}, {Anekallu}, {Arciuli}, {Arnett}, {Ascolese}, {Bancroft}, {Bland}, {Brysch}, {Calvanese}, {Castronuovo}, {{\v{C}}erm{\'a}k}, {Chornay}, {Clemens}, {Coker}, {Collinson}, {D'Amicis}, {Dandouras}, {Darnley}, {Davies}, {Davison}, {De Los Santos}, {Devoto}, {Dirks}, {Edlund}, {Fazakerley}, {Ferris}, {Frost}, {Fruit}, {Garat}, {G{\'e}not}, {Gibson}, {Gilbert}, {de Giosa}, {Gradone}, {Hailey},
  {Horbury}, {Hunt}, {Jacquey}, {Johnson}, {Lavraud}, {Lawrenson}, {Leblanc}, {Lockhart}, {Maksimovic}, {Malpus}, {Marcucci}, {Mazelle}, {Monti}, {Myers}, {Nguyen}, {Rodriguez-Pacheco}, {Phillips}, {Popecki}, {Rees}, {Rogacki}, {Ruane}, {Rust}, {Salatti}, {Sauvaud}, {Stakhiv}, {Stange}, {Stubbs}, {Taylor}, {Techer}, {Terrier}, {Thibodeaux}, {Urdiales}, {Varsani}, {Walsh}, {Watson}, {Wheeler}, {Willis}, {Wimmer-Schweingruber}, {Winter}, {Yardley}, \& {Zouganelis}}]{owen2020solar}
{Owen}, C.~J., {Bruno}, R., {Livi}, S., {et~al.} 2020, \bibinfo{title}{{The Solar Orbiter Solar Wind Analyser (SWA) suite},} \aap, 642, A16, \dodoi{10.1051/0004-6361/201937259}

\bibitem[{E. {Palmerio} {et~al.}(2017){Palmerio}, {Kilpua}, {James}, {Green}, {Pomoell}, {Isavnin}, \& {Valori}}]{2017Palmerio}
{Palmerio}, E., {Kilpua}, E.~K.~J., {James}, A.~W., {et~al.} 2017, \bibinfo{title}{{Determining the Intrinsic CME Flux Rope Type Using Remote-sensing Solar Disk Observations},} \solphys, 292, 39, \dodoi{10.1007/s11207-017-1063-x}

\bibitem[{S. {Patsourakos} {et~al.}(2020){Patsourakos}, {Vourlidas}, {T{\"o}r{\"o}k}, {Kliem}, {Antiochos}, {Archontis}, {Aulanier}, {Cheng}, {Chintzoglou}, {Georgoulis}, {Green}, {Leake}, {Moore}, {Nindos}, {Syntelis}, {Yardley}, {Yurchyshyn}, \& {Zhang}}]{2020Patsourakos}
{Patsourakos}, S., {Vourlidas}, A., {T{\"o}r{\"o}k}, T., {et~al.} 2020, \bibinfo{title}{{Decoding the Pre-Eruptive Magnetic Field Configurations of Coronal Mass Ejections},} \ssr, 216, 131, \dodoi{10.1007/s11214-020-00757-9}

\bibitem[{W.~D. {Pesnell} {et~al.}(2012){Pesnell}, {Thompson}, \& {Chamberlin}}]{pesnell_2012}
{Pesnell}, W.~D., {Thompson}, B.~J., \& {Chamberlin}, P.~C. 2012, \bibinfo{title}{{The Solar Dynamics Observatory (SDO)},} \solphys, 275, 3, \dodoi{10.1007/s11207-011-9841-3}

\bibitem[{D.~V. {Reames}(1999){Reames}}]{reames_1999}
{Reames}, D.~V. 1999, \bibinfo{title}{{Particle acceleration at the Sun and in the heliosphere},} \ssr, 90, 413, \dodoi{10.1023/A:1005105831781}

\bibitem[{D.~V. {Reames}(2013){Reames}}]{reames_2013}
{Reames}, D.~V. 2013, \bibinfo{title}{{The Two Sources of Solar Energetic Particles},} \ssr, 175, 53, \dodoi{10.1007/s11214-013-9958-9}

\bibitem[{P. {Riley}(2012){Riley}}]{riley_2012}
{Riley}, P. 2012, \bibinfo{title}{{On the probability of occurrence of extreme space weather events},} Space Weather, 10, 02012, \dodoi{10.1029/2011SW000734}

\bibitem[{A. {Sahade} {et~al.}(2026){Sahade}, {Mac Cormack}, {Vourlidas}, {Nieves-Chinchilla}, {Downs}, {Sasso}, \& {Karpen}}]{sahade_2026}
{Sahade}, A., {Mac Cormack}, C., {Vourlidas}, A., {et~al.} 2026, \bibinfo{title}{{Understanding the Complex Morphology of a Coronal Mass Ejection. II. How Pre-eruptive Conditions Shape Coronal Mass Ejection Evolution},} \apj, 1003, 154, \dodoi{10.3847/1538-4357/ae64f9}

\bibitem[{P.~H. {Scherrer} {et~al.}(2012){Scherrer}, {Schou}, {Bush}, {Kosovichev}, {Bogart}, {Hoeksema}, {Liu}, {Duvall}, {Zhao}, {Title}, {Schrijver}, {Tarbell}, \& {Tomczyk}}]{scherrer_2012}
{Scherrer}, P.~H., {Schou}, J., {Bush}, R.~I., {et~al.} 2012, \bibinfo{title}{{The Helioseismic and Magnetic Imager (HMI) Investigation for the Solar Dynamics Observatory (SDO)},} \solphys, 275, 207, \dodoi{10.1007/s11207-011-9834-2}

\bibitem[{C.~J. {Schrijver} {et~al.}(2012){Schrijver}, {Beer}, {Baltensperger}, {Cliver}, {G{\"u}del}, {Hudson}, {McCracken}, {Osten}, {Peter}, {Soderblom}, {Usoskin}, \& {Wolff}}]{schrijver_2012}
{Schrijver}, C.~J., {Beer}, J., {Baltensperger}, U., {et~al.} 2012, \bibinfo{title}{{Estimating the frequency of extremely energetic solar events, based on solar, stellar, lunar, and terrestrial records},} Journal of Geophysical Research (Space Physics), 117, A08103, \dodoi{10.1029/2012JA017706}

\bibitem[{C.~J. {Schrijver} {et~al.}(2015){Schrijver}, {Kauristie}, {Aylward}, {Denardini}, {Gibson}, {Glover}, {Gopalswamy}, {Grande}, {Hapgood}, {Heynderickx}, {Jakowski}, {Kalegaev}, {Lapenta}, {Linker}, {Liu}, {Mandrini}, {Mann}, {Nagatsuma}, {Nandy}, {Obara}, {Paul O'Brien}, {Onsager}, {Opgenoorth}, {Terkildsen}, {Valladares}, \& {Vilmer}}]{2015Schrijver}
{Schrijver}, C.~J., {Kauristie}, K., {Aylward}, A.~D., {et~al.} 2015, \bibinfo{title}{{Understanding space weather to shield society: A global road map for 2015-2025 commissioned by COSPAR and ILWS},} Advances in Space Research, 55, 2745, \dodoi{10.1016/j.asr.2015.03.023}

\bibitem[{R. {Schwenn}(2006){Schwenn}}]{2006Schwenn}
{Schwenn}, R. 2006, \bibinfo{title}{{Space Weather: The Solar Perspective},} Living Reviews in Solar Physics, 3, 2, \dodoi{10.12942/lrsp-2006-2}

\bibitem[{A.~K. {Singh} {et~al.}(2021){Singh}, {Bhargawa}, {Siingh}, \& {Singh}}]{2021Singh}
{Singh}, A.~K., {Bhargawa}, A., {Siingh}, D., \& {Singh}, R.~P. 2021, \bibinfo{title}{{Physics of Space Weather Phenomena: A Review},} Geosciences, 11, 286, \dodoi{10.3390/geosciences11070286}

\bibitem[{M. Temmer {et~al.}(2023)Temmer, Scolini, Richardson, Heinemann, Paouris, Vourlidas, Bisi, {Al-Haddad}, Amerstorfer, Barnard, Bure{\v s}ov{\'a}, Hofmeister, Iwai, Jackson, Jarolim, Jian, Linker, Lugaz, Manoharan, Mays, Mishra, Owens, Palmerio, Perri, Pomoell, Pinto, Samara, Singh, Sur, Verbeke, Veronig, \& Zhuang}]{2023Temmer}
Temmer, M., Scolini, C., Richardson, I.~G., {et~al.} 2023, \bibinfo{title}{{{CME}} Propagation through the Heliosphere: {{Status}} and Future of Observations and Model Development,} AdSpR, in press, \dodoi{10.1016/j.asr.2023.07.003}

\bibitem[{A.~F. {Thernisien} {et~al.}(2025){Thernisien}, {Chua}, {Carter}, {Rich}, {Noya}, {Babich}, {Crippa}, {Baugh}, {Bordlemay}, {Socker}, {Biesecker}, {Korendyke}, {Wang}, {Vassiliadis}, {Wang}, {Abbay}, {Bagnall}, {Balmaceda}, {Brown}, {Bonafede}, {Boyer}, {Declet}, {Cheng}, {Corsi}, {Cremerius}, {Chavis}, {Chiralo}, {Clifford}, {Dancheck}, {Davis}, {Dima}, {Dudley}, {Gardner}, {Gardner}, {Hagood}, {Hagood}, {Hohl}, {Hunt}, {Jenkins}, {Johnson}, {Koehler}, {Kuroda}, {Lanagan}, {Laut}, {Lynch}, {Mallory}, {Mechel}, {Miles}, {Miranda}, {Newman}, {Nguyen}, {Ogindo}, {Pellak}, {Podgurski}, {Ragan}, {Richards}, {Silver}, {Simmons}, {Schmit}, {Smith}, {Spitzak}, {Tadikonda}, {Tanner}, {Uhl}, {Verzosa}, {Walker}, {Wiggins}, {Williams}, {Wilson}, \& {Zurcher}}]{thernisien_2025}
{Thernisien}, A.~F., {Chua}, D.~H., {Carter}, M.~T., {et~al.} 2025, \bibinfo{title}{{The CCOR Compact Coronagraphs for the Geostationary Operational Environmental Satellite-19 (GOES-19) and the Space Weather Follow On (SWFO) Missions},} arXiv e-prints, arXiv:2508.13467, \dodoi{10.48550/arXiv.2508.13467}

\bibitem[{V.~S. {Titov} {et~al.}(2018){Titov}, {Downs}, {Miki{\'c}}, {T{\"o}r{\"o}k}, {Linker}, \& {Caplan}}]{titov_2018}
{Titov}, V.~S., {Downs}, C., {Miki{\'c}}, Z., {et~al.} 2018, \bibinfo{title}{{Regularized Biot-Savart Laws for Modeling Magnetic Flux Ropes},} \apjl, 852, L21, \dodoi{10.3847/2041-8213/aaa3da}

\bibitem[{T. {T{\"o}r{\"o}k} {et~al.}(2018){T{\"o}r{\"o}k}, {Downs}, {Linker}, {Lionello}, {Titov}, {Miki{\'c}}, {Riley}, {Caplan}, \& {Wijaya}}]{torok_2018}
{T{\"o}r{\"o}k}, T., {Downs}, C., {Linker}, J.~A., {et~al.} 2018, \bibinfo{title}{{Sun-to-Earth MHD Simulation of the 2000 July 14 {\textquotedblleft}Bastille Day{\textquotedblright} Eruption},} \apj, 856, 75, \dodoi{10.3847/1538-4357/aab36d}

\bibitem[{B.~T. {Tsurutani} {et~al.}(2006){Tsurutani}, {Gonzalez}, {Gonzalez}, {Guarnieri}, {Gopalswamy}, {Grande}, {Kamide}, {Kasahara}, {Lu}, {Mann}, {McPherron}, {Soraas}, \& {Vasyliunas}}]{2006Tsurutani}
{Tsurutani}, B.~T., {Gonzalez}, W.~D., {Gonzalez}, A. L.~C., {et~al.} 2006, \bibinfo{title}{{Corotating solar wind streams and recurrent geomagnetic activity: A review},} Journal of Geophysical Research (Space Physics), 111, A07S01, \dodoi{10.1029/2005JA011273}

\bibitem[{D.~F. {Webb} \& T.~A. {Howard}(2012){Webb} \& {Howard}}]{webb_2012}
{Webb}, D.~F., \& {Howard}, T.~A. 2012, \bibinfo{title}{{Coronal Mass Ejections: Observations},} Living Reviews in Solar Physics, 9, 3, \dodoi{10.12942/lrsp-2012-3}

\bibitem[{Z. {Zhou} {et~al.}(2020){Zhou}, {Liu}, {Cheng}, {Jiang}, {Wang}, {Liu}, \& {Cui}}]{2020Zhou}
{Zhou}, Z., {Liu}, R., {Cheng}, X., {et~al.} 2020, \bibinfo{title}{{The Relationship between Chirality, Sense of Rotation, and Hemispheric Preference of Solar Eruptive Filaments},} \apj, 891, 180, \dodoi{10.3847/1538-4357/ab7666}

\end{thebibliography}
\bibliographystyle{aasjournalv7}

%
%
%
%
%

\end{document}